\documentclass[a4paper,11pt]{article}
\pdfoutput=1 % if your are submitting a pdflatex (i.e. if you have
\usepackage{jcappub} % for details on the use of the package, please
\usepackage{graphicx}
\usepackage{amsmath}
\usepackage{amssymb}
\usepackage[all]{hypcap}
\usepackage{mathptmx}
\usepackage{txfonts}
\usepackage{ae,aecompl}
\usepackage[normalem]{ulem}
\usepackage{longtable}
\usepackage{multirow}
\usepackage{multicol}
\usepackage{tablefootnote}
\usepackage{appendix}
\usepackage{xcolor}
\usepackage{pagecolor}
\usepackage{hyperref}
\usepackage{chngcntr}

\newcommand{\hmpc}{{\, h^{-1}\, {\rm Mpc}}}

\definecolor{ss}{RGB}{210, 60, 0}
\definecolor{ab}{RGB}{250, 0, 120}
\definecolor{gn}{RGB}{0, 90, 190}

\title{Analyzing the cosmic web environment in the vicinity of grand-design and flocculent spirals with local geometric index}

\author[a, b]{Suman Sarkar}
\author[a]{, Ganesh Narayanan}
\author[a]{, and Arunima Banerjee}

\affiliation[a]{Department of Physics, Indian Institute of Science Education and Research Tirupati,\\ Tirupati - 517507, India.}
\affiliation[b]{Department of Physics, Indian Institute of Technology Kharagpur,\\ West Bengal - 721302, India.}

\emailAdd{arunima@iisertirupati.ac.in}
\abstract{We explore the environment of a combined set of $367$ grand-design and $619$ flocculent spiral galaxies. We introduce a novel estimator called the \textit{local geometric index} to quantify the morphology of the local environment of these $986$ spirals. The local geometric index allows us to classify the environment of galaxies into voids, sheets, filaments, and clusters. We find that grand-designs are mostly located in dense environments like clusters and filaments ($\sim 78\%$), whereas the fraction of the flocculents lying in sparse environments like voids and sheets is significantly higher ($ > 10\%$) than that of the grand-designs. A $p$-value $<$ $10 ^{-3}$ from a Kolmogorov-Smirnov test indicates that our results are statistically significant at $99.9\%$ confidence level. Further, we note that dense environments with large tidal flows are dominated by the grand-designs. On the other hand, low-density environments such as sheets and voids favor the growth of flocculents.}


\begin{document}
\maketitle
\flushbottom

% \begin{keywords}
% galaxies: galaxy dynamics - galaxy morphology - galaxy evolution - Large scale structure of the universe: cosmic web - redshift surveys.
% \end{keywords}

\section{Introduction}
Spiral galaxies are ubiquitous. In fact, more than $ 60 \%$ of the galaxies in the nearby universe are spirals \cite{Loveday_1996, Willett2013}. The spiral arms possibly constitute the most prominent non-axisymmetric features observed in disc galaxies. Being the overdense regions in the galactic disc, they favour the compression of gas clouds and hence constitute the primary sites of enhanced star formation \cite{Elmegreen2011SF, Yu_2021SF}. Furthermore, the quadrupole moment associated with the gravitational potential of non-axisymmetric disc dynamical features, such as the bars and spiral arms, generates torques that aid in the transfer of angular momentum in the disc \cite{Binney2008}. Spirals also induce gas inflow from the outer regions of the galaxy, which feeds the central active galactic nuclei \cite{Ho1997AGN, Hunt1999AGN}. Thus, the spiral arms play a vital role in the general evolution of the galactic center and the disc \cite{Querejeta2016, Yu_2022}. However, the emergence and growth of the spiral features in disc galaxies is not completely understood yet, nor is the role of the environment in which they grow. \cite{Dobbs2014}. \\

In general, the morphological classification of disc galaxies is based on the presence or absence of a bar as well as the degree of winding of the spiral arms \cite{Hubble1926,de_Vaucou1959H}. A 12-point classification scheme was introduced based on the number of spiral arms and their arm contrast by \cite{elmegreen82, elmegreen87}. The continuous and well-defined spirals were termed as \textit{grand-design} spirals, whereas the patchy and irregular ones were referred to as \textit{flocculents}. The observed morphology of the spiral arm is mostly regulated by its formation mechanism \cite{Dobbs2014}. The flocculents can be looked upon as local spiral features generated the local non-axisymmetric instabilities, which get amplified by swing amplification \cite {goldreich65a, Toomre1981}. The density wave theory models the formation of a stationary spiral structures like grand-designs \cite{lin64}, which, among others, can be triggered by a bar-like instability \cite{kormendy79}. Interestingly, N-body simulations fail to generate stationary global spiral patterns \cite{Sellwood2011} and instead give rise to transient global spiral features. This seems plausible due to the superposition of several normal modes \cite{Sellwood_Carlberg2014}. The large scale flocculent features could also be induced by stochastic, self-propagating star formation model where stars are born from shock waves triggered by high mass stars. This propagates as a chain reaction and further gets sheared by differential rotation of the galaxy\cite{Gerola1978}. Therefore, the internal mechanisms responsible for spiral formation are quite complex and  could be influenced by external drivers.\\

\begin{figure*}
\resizebox{7.5 cm}{!}{\rotatebox{0}{\includegraphics{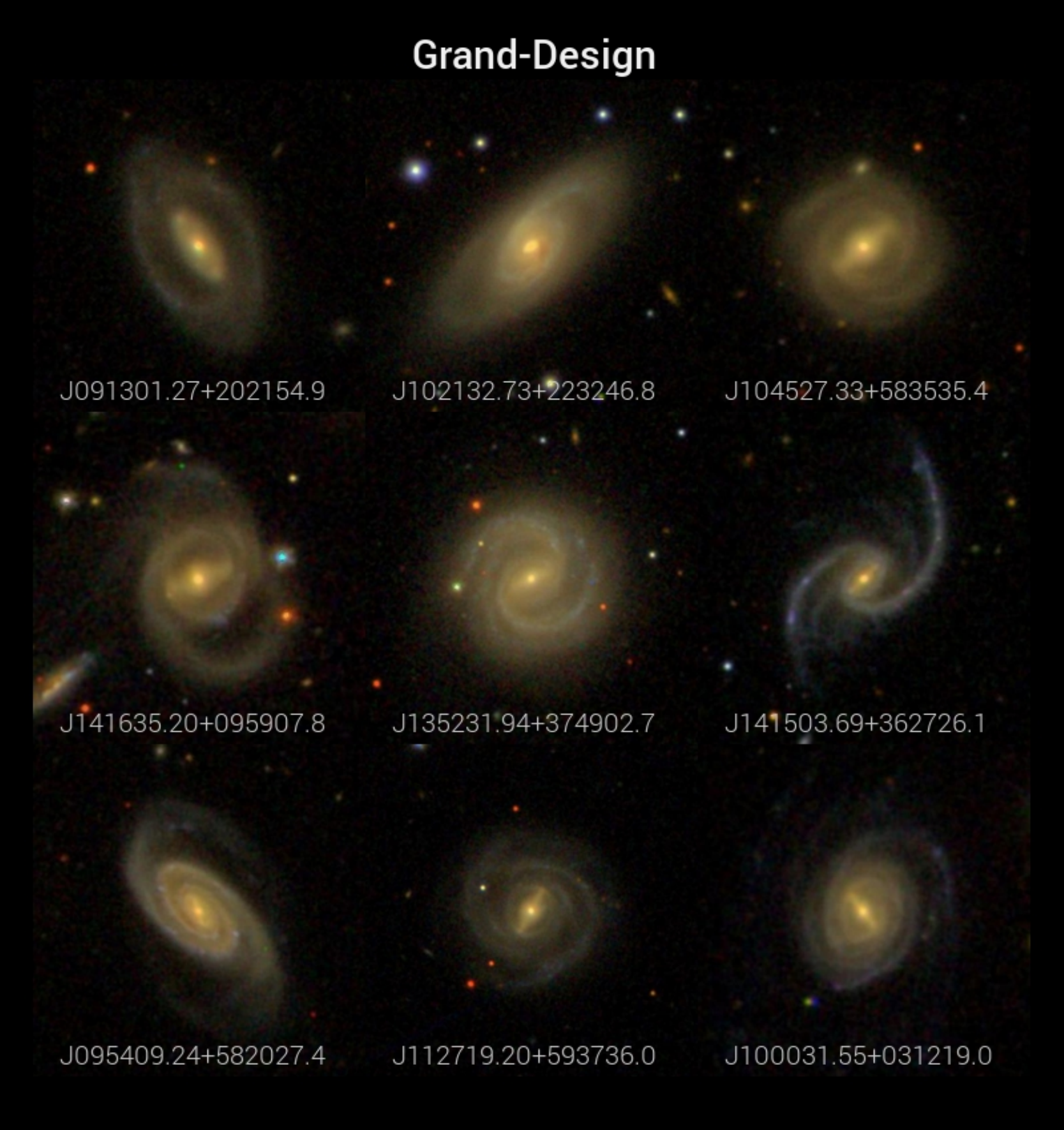}}} \hspace{0.2 cm}
\resizebox{7.5 cm}{!}{\rotatebox{0}{\includegraphics{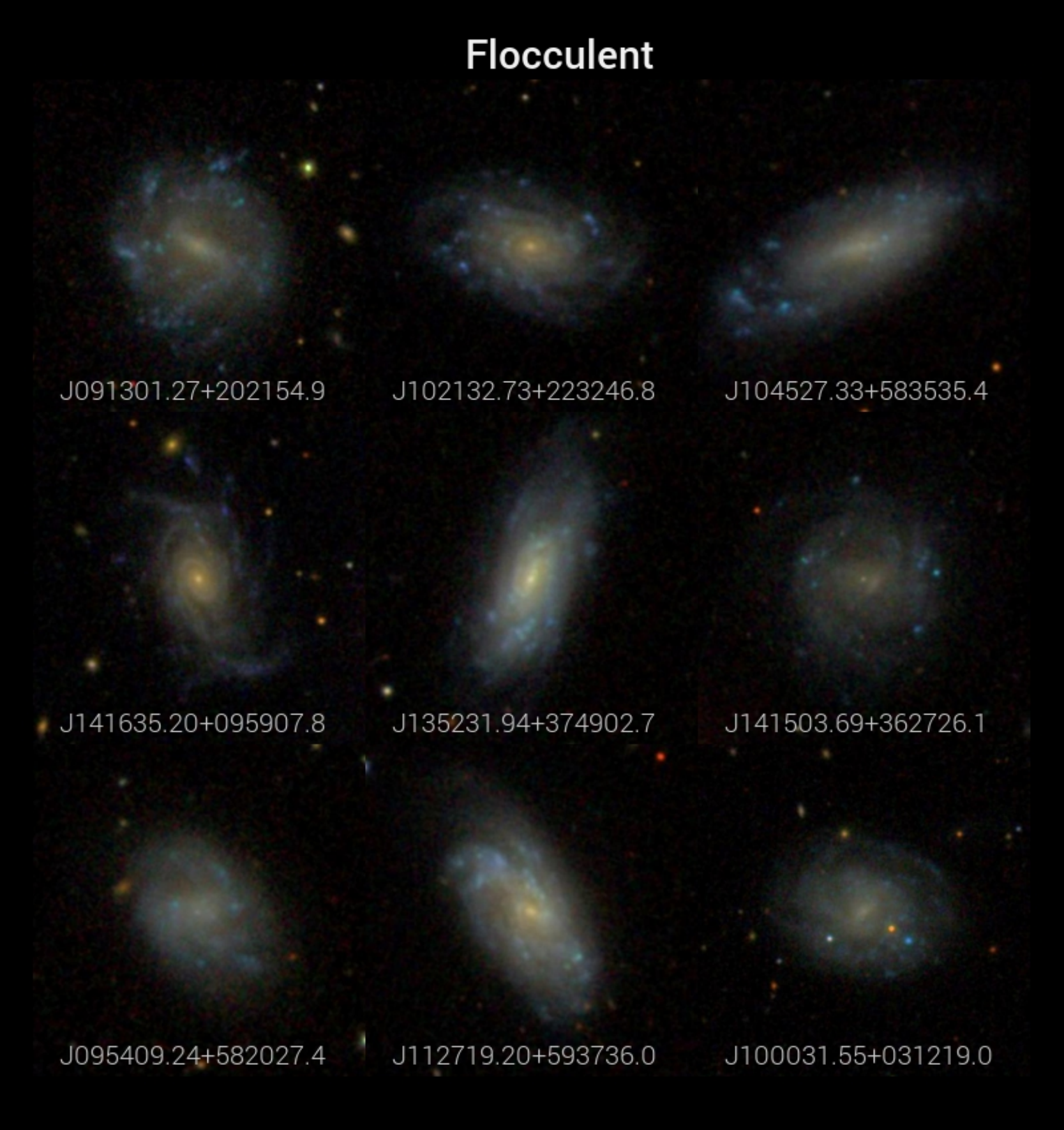}}}
\label{fig:sdssclass}
\caption{A montage of grand-design  \textbf{[Left]} and  flocculent \textbf{ [Right]} spiral galaxies from the dataset used.}
\end{figure*}

There have been several attempts in the past on studying the effects of the environment on galaxy morphology, both at small and large (cosmological) scales \cite{oemler74,davis76,dressler80,hogg03,balogh04,kauffmann04,bamform09,peng10,skibba12,pandey17,pandey20,sarkar21}. It is now well known that spirals could be triggered by the tidal field of an interacting galaxy or that of a cluster in which a galaxy is hosted \cite{holmberg41,toomre72}. Using N-body simulations, \cite{Byrd1992} finds that prograde companions can trigger grand-design patterns in host galaxies. Here the tidal perturbation was found to be inversely proportional to the cube of the distance of closest approach, and proportional to the product of the cube of disc size and the mass ratio of the perturber and the host. Furthermore, they noted that perturbers which fail to excite grand-design patterns at least produce filamentary spiral features. N-body + hydrodynamical simulations by \cite{Pettitt2016} show the spirals in the stellar and the gaseous discs induced by tidal flybys exhibit little difference in arm morphology, pitch angles and pattern speeds. \cite{Dobbs2010} model the spiral in M51 as tidally-induced by its companion NGC 5195. Similarly, a simulation study by \cite{Semczuk2018} models the observed two-armed spiral structure in M33 from the tidal interaction with M31. Finally, by considering a sample of spiral galaxies from the Spitzer Infrared Nearby Galaxies Survey (SINGS), \cite{Kendall_2011} correlates the existence of strong grand-design spirals with the presence of a companion galaxy in general. However, there may be galaxies with grand-design spirals without interacting  companions; these could be in the tidal-field of dark sub-halos, which are not visible \cite{Sellwood2022}. \cite{Semczuk_2017}, on the other hand, finds that the spiral survives longer if it has extended orbits in the potential of a galaxy cluster. Besides, they observe that nine out of the twenty-four grand-designs in the Virgo cluster do not show signs of recent interactions, indicating the role of the net tidal field of the cluster in exciting spiral instabilities. Further, studying a sample of galaxies from ILLUSTRIS-TNG100, \cite{Lokas2020} shows that the tidal field of the cluster may induce non-axi-symmetric features and enhance the strength of the same. Hence, the overall understanding of spiral activity in galaxies on the basis of their environment is still fuzzy and needs to be explored further, particularly in the context of the cosmic web's environment, which has received very little attention. \\

The large-scale tidal forces cause anisotropic collapse \cite{guyot70, Peebles81} of large-scale structures, forming voids, sheets, filaments and clusters, and hence the cosmic web. So far, a number of methods have been developed to identify the geometry of the different structures in the cosmic web, including the percolation analysis \cite{zeldovich82,shandarin89,shandarin10,aragon10b}, the minimal spanning tree \cite{barrow85,graham95,colberg07}, the Minkowski functionals \cite{mecke94,sarkar22}, Delaunay tessellation field estimator (DTFE) \cite{schaap00,waygaert09}, manifolds in the tidal field \cite{hahn07a,hahn07b,forero09,cautun12}, Lagrangian sub-manifolds \cite{ramachandra15}, local dimension \cite{sarkar09,sarkar19}, Bayesian sampling algorithm \cite{jasche10}, skeleton analysis \cite{novikov06,sousbie08}, discrete Morse theory \cite{sousbie11}, phase space folds \cite{flack12} and machine learning \cite{buncher20}. Different components of the cosmic web are also studied separately in a number of studies \cite{sousbie08b,aragon10,alpaslan14,sutter14,tempel14,cohn22}. To identify the geometric environments around galaxies, we present a novel estimator in this study called the {\it local geometric index}. The local geometric index is a  simple yet effective tool for classifying the various elements of the cosmic web. It also offers a few advantages over some of the other estimators, such as local density and local dimension. The local geometric index is useful to classify the galaxies in different components of the cosmic web, especially in sparse galaxy distributions mapped by redshift surveys. It works without analyzing the underlying density field continuum or identifying precise boundary-components or critical points of the 3-manifold. Therefore, this method offers a simple yet effective scheme for the classification of cosmic web environments. We determine the local geometric index of $986$ spiral galaxies classified into grand-designs and flocculents \cite{buta15, gdfl2022} \\

The rest of the paper is organized as follows. In \S 2, we present the data to be analysed. \S 2.1 describes the sample of the grand-designs and the flocculents whose local geometric indices are to be determined. In \S 2.2, we present the data used as the background galaxy distribution. In \S 3, we discuss the methods of analysis, defining the local geometric index in \S 3.3 and the Kolmogorov-Smirnov (KS) test that we carry out is discussed in \S 3.4. In \S 4, we present the results, followed by a discussion in \S 5 and the conclusions in \S 6.
%##############################################     DATA   #############################################################
\begin{figure*}
\hspace{-15 px}
\resizebox{16 cm}{!}{\rotatebox{0}{\includegraphics{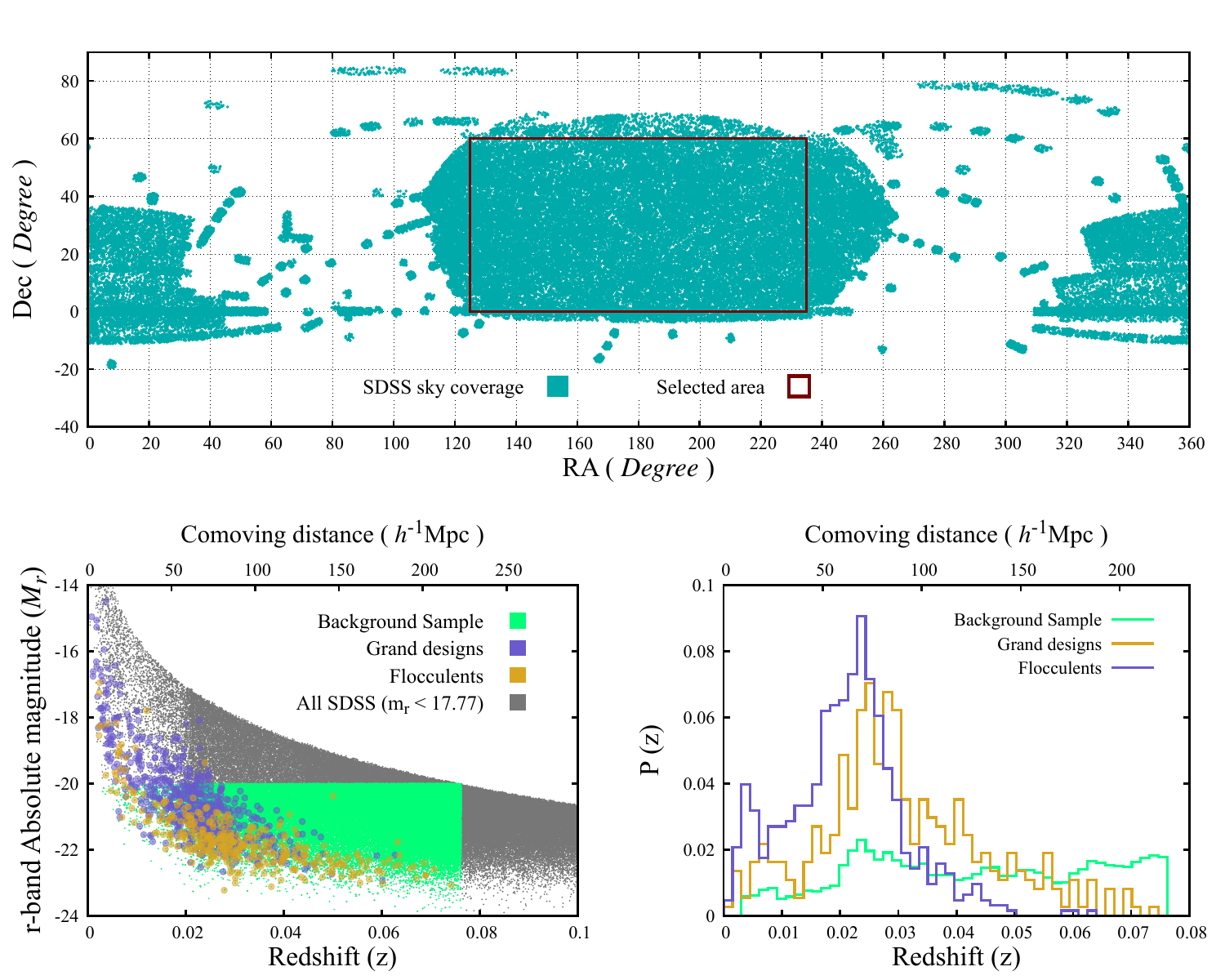}}} 
\label{fig:sample}
\caption{\textbf{[Top]} The sky coverage of the selected sample. \textbf{[Bottom-left]} The redshift-Absolute magnitude distribution of grand-design and flocculent galaxies is shown along with the background distribution. \textbf{[Bottom-right]} The probability distribution of the galaxies as a function of redshift.}
\end{figure*}

\section{Data}
\subsection{Sample of grand-design and flocculent spirals}
We study the geometry of the local environment of a number of grand-designs and flocculents from the seventeenth data release of the Sloan Digital Sky Survey (SDSS DR17) \cite{york00,abdurrouf22}. To start with, we consider a combined set of $1975$ grand-designs and flocculents. Among these, $201$ grand-designs and $554$ flocculents are identified by \cite{buta15} using the observed arm-interarm contrast of these spirals. The remaining $499$ grand-designs and $721$ flocculents are identified by \cite{gdfl2022} from the SDSS. These galaxies are classified using a deep convolutional neural network (DCNN) comprising five layers of convolution, three max-pooling layers, and two dense layers. The DCNN model is trained on the images of galaxies classified by \cite{buta15} and further predicts the morphological class of the $1220$ spiral galaxies with $97\%$ average accuracy. Out of these $1220$, $252$ grand-designs and $604$ flocculents were classified with $100\%$ confidence; the rest with a confidence level of $80-100\%$. A montage of some of these grand-design and flocculent spirals is presented in \autoref{fig:sdssclass}. Here, we only consider the galaxies from the combined set which lie within the equatorial coordinate range $125^\circ \leq \mathrm{RA} \leq 235^\circ$ and $0^\circ \leq \mathrm{Dec} \leq 60^\circ$ where we find the SDSS sky coverage to be contiguous, i.e., without holes and patches (Top panel of \autoref{fig:sample}). Also, to obtain the geometry of the background galaxy distribution (\autoref{sec:bgdata}) we  restrict the redshift range to $ z \leq 0.0762$. Using the spectroscopic redshift and equatorial coordinates, we determine the co-moving coordinates of these galaxies. Further, to investigate the dependence of the different physical properties of the grand-design and the flocculent spiral galaxies on the geometry of their local environment, the asymptotic rotational velocity ($V_{\rm{rot}}$) and total atomic hydrogen mass ($M_{HI}$) of the sample galaxies are obtained from Hyperleda\footnote{http://leda.univ-lyon1.fr/} \cite{makarov14}. The stellar mass ($M_{\star}$), specific star formation rate ($sSFR$) and age of the stellar population ($\tau_{\star}$) of these galaxies are taken from the table \textit{stellarMassFSPSGranWideDust} in the SDSS DR17 database. This table presents the physical properties of the galaxies mapped using the Granada stellar population synthesis model \cite{conroy09,conroy10} to fit SDSS photometry. We finally have $370$ grand-designs and $629$ flocculents for which we have all the required information.

\subsection{Data for the environment: The background galaxy distribution }
\label{sec:bgdata}
The distribution of our sample grand-designs and flocculents is very sparse and also inhomogeneous in nature due to Malmquist bias \cite{malmquist25}, i.e. intrinsically faint galaxies missing at higher redshifts (bottom left panel of \autoref{fig:sample}). Instead of using a selection function for counterbalancing this selection bias, we use a magnitude-limited sample (bottom left panel of \autoref{fig:sample}) to map the neighbourhood of the grand-designs and flocculents. A sample with an upper bound of absolute magnitude ensures the galaxies with similar intrinsic brightness are accessible across the entire redshift range probed. It is important to have the number density profile for the background sample free from the selection biases of the survey (bottom right panel of \autoref{fig:sample}). This is to ensure that the variation in number density while mapping the neighbourhood of any given grand-design or flocculent is only due to the fluctuations in the actual matter distribution of the Universe. To prepare the background galaxy sample, we use data from SDSS DR17. The spectroscopic redshifts are used to obtain the exact co-moving coordinates of the galaxies in 3D. A uniform region within $125^\circ \leq \mathrm{RA} \leq 235^\circ$ and $0^\circ \leq \mathrm{Dec} \leq 60^\circ$ is selected from the entire SDSS sky coverage that matches the sample of grand-designs and flocculents. The magnitude-limited sample we prepare within this angular span has {\it r}-band Petrosian absolute magnitude limit $M_r \leq -20$, which leads to the redshift range $ z \leq 0.0762$. Finally, we have $90,882$ galaxies in the background galaxy distribution that meet the aforesaid conditions. The distribution have a mean number density ($\bar{\rho}$) of $ 1.45 \times 10^{-2} \,h^{3}\,\rm{Mpc}^{-3}$. Comoving distances for all the galaxies used in this study are evaluated by considering $\Lambda CDM$ cosmology with $H_0 = 100\;h = 67.4 \;\mathrm{Km\;s^{-1}\; Mpc^{-1}}$, $\Omega_{\Lambda}=0.685$ and $\Omega_{m}=0.315$ \cite{planck2018}. \\

% intergalactic separation of $4.64 \hmpc$ 

%######################################     Methods of analysis  #######################################################

\section{Method of Analysis}
\subsection{Local dimension}
\textbf{We start off with the definition of the local dimension ($D$) \cite{sarkar09,sarkar19} of a galaxy}. The 3-D comoving, cartesian coordinates of the galaxy in consideration are estimated first, using their spectroscopic redshifts and equatorial coordinates. A comoving sphere of an arbitrary radius $R$ is considered around the galaxy, and the number of neighbouring galaxies inside the sphere $N$ is noted. This exercise is then repeated for a number of values of $R$ up to a certain radius $R_{max}$. If the density profile and geometrical nature of the local environment does not change up to $R_{max}$, then in principle the relation
\begin{eqnarray}
 N(< R)= C R^{D}
\label{eq:ld1}
\end{eqnarray}
\noindent should hold good  within that range. Here $D$ is the local dimension of the cosmic web around that galaxy and $C$ acts like a proportionality constant for minimal variation in density contrast within the specified radial span. Here, $D \sim 1$ would represent one-dimensional structures, such as filaments. $D \sim 2$ suggests that the galaxy is associated with a two-dimensional system, such as a wall or sheet. Furthermore, galaxies with $D \sim 3$ are the ones in a locally isotropic environment \cite{sarkar09,sarkar19}. In each of these situations, The value of $C$ would be different depending on the geometry and density of the local environment. \\

\noindent Taking logarithm of \autoref{eq:ld1} gives
\begin{eqnarray}
\psi = \gamma + D \phi,
\label{eq:ld2}
\end{eqnarray}

\noindent where $\psi=\log{N}$, $\gamma=\log{C}$  and $\phi = \log{R}$. We vary $R$ with a uniform increment of $\Delta R=0.1 \times R_{max}$. For different values of $R$, we find different observed values of $\psi$ and $\phi$ in \autoref{eq:ld2}. 
 The local dimension $D$ for the galaxy is then estimated by performing the least square fit to all these values of $(\psi, \, \, \phi)$ within the range $0 < R \leq R_{max}$. With a degree of freedom $\nu=9$ (for 10 data points). The chi-square is calculated as 

\begin{eqnarray}
\chi^{2} = \sum { \left( \frac{  \psi -  \overline{\psi} } {\Delta\psi}\right)^2},
\label{eq:chisqr}
\end{eqnarray}

\noindent where $\overline{\psi}$ is obtained from the best-fitted values of $\gamma$ and $D$, for a galaxy whose environment is being studied. $\Delta \psi$ is calculated using the quantities $\Delta\gamma$ and $\Delta D$, obtained from the least square fit, and is given by \textbf{$\Delta\psi = \sqrt{\Delta\gamma^2 + \phi^2 \Delta D^2}$}. We set a criterion of $\frac{\chi^2}{\nu} \leq \frac{1}{2}$ in order to retain only those galaxies which have constant $D$ and $\gamma$ throughout the considered length scale $R_{max}$. We note that this filtering neither removes many galaxies nor biases the sample.\\

\subsubsection{Determination of $R_{max}$ for a particular galaxy}
One question that may arise here is: \textit{What is the maximum length scale that can be considered ``local"?} Here, the choice of $R_{max}$ plays a crucial role. A galaxy's geometric environment can be characterized differently for different values of $R_{max}$. A galaxy can be part of a filament, showing $D \sim 1$ for a small value of $R_{max}$. Now, if we gradually increase $R_{max}$, the same galaxy can be found to be part of a sheet, i.e., $D \sim 2$. In general, for small values of $R_{max}$, galaxies with $D \sim 3$ can correspond to either cluster cores or fields. Now, if $R_{max}$ is chosen to be larger than the size of the cluster, we get $D \sim 1.5$ for these galaxies, as they lie at the junction of filaments. However, the field galaxies will still have $D \sim 3$, despite the increment in $R_{max}$. Taking a very large value of $R_{max}$ one can eventually reach up to the scale of homogeneity \cite{sarkar19}, where $D \sim 3$ is obtained for each galaxy. The size of a galaxy cluster can vary from a few mega-parsecs to tens of mega-parsecs \cite{smith12,wen22}. Filaments and sheets can also have largely varying sizes \cite{pandey11,tempel14,duque22}. In the earlier studies, fixed values of $R_{max}$ were used to find the local dimension. In this work, we refrain from using a constant value of $R_{max}$ for all the galaxies and keep the choice free instead. The density profile around the galaxy is checked for increasing $R$. As we use a spherical filter of variable size to count the neighbouring galaxies, any fundamental geometrical change in the environment is also expected to show its signature through the variation in density. We identify a length scale $R_{max}$ up to which there is no structural change in the cosmic web environment. This allows us to have an adaptive spherical filter for local dimension estimation. Using the mean number density ($\bar{\rho}$) of the background galaxy distribution (\autoref{sec:bgdata}), one can define the density contrast at radius $R$ as

\begin{eqnarray}
\delta(R)=\frac{3N}{4\pi \bar{\rho} R^3} -1 .
\label{eq:density}
\end{eqnarray}

\noindent  $\delta(R)$ is calculated for different values of $R$ starting from the mean intergalactic separation. By increasing values of $R$, we identify the range where $\delta(R)$ reaches the minimum variation. The minimum, maximum and median of $\delta (R)$ are noted  to find the normalized absolute deviation in density contrast at $R$ as

\begin{eqnarray}
\eta (R) = \left \lvert \frac{\delta (R) -\delta_{\rm{median}}}{\delta_{\rm{max}}-\delta_{\rm{min}}} \right \rvert.
\label{eq:lden}
\end{eqnarray}

\begin{figure*}
\centering
\vspace{-10 px}
\resizebox{16 cm}{!}{\rotatebox{0}{\includegraphics{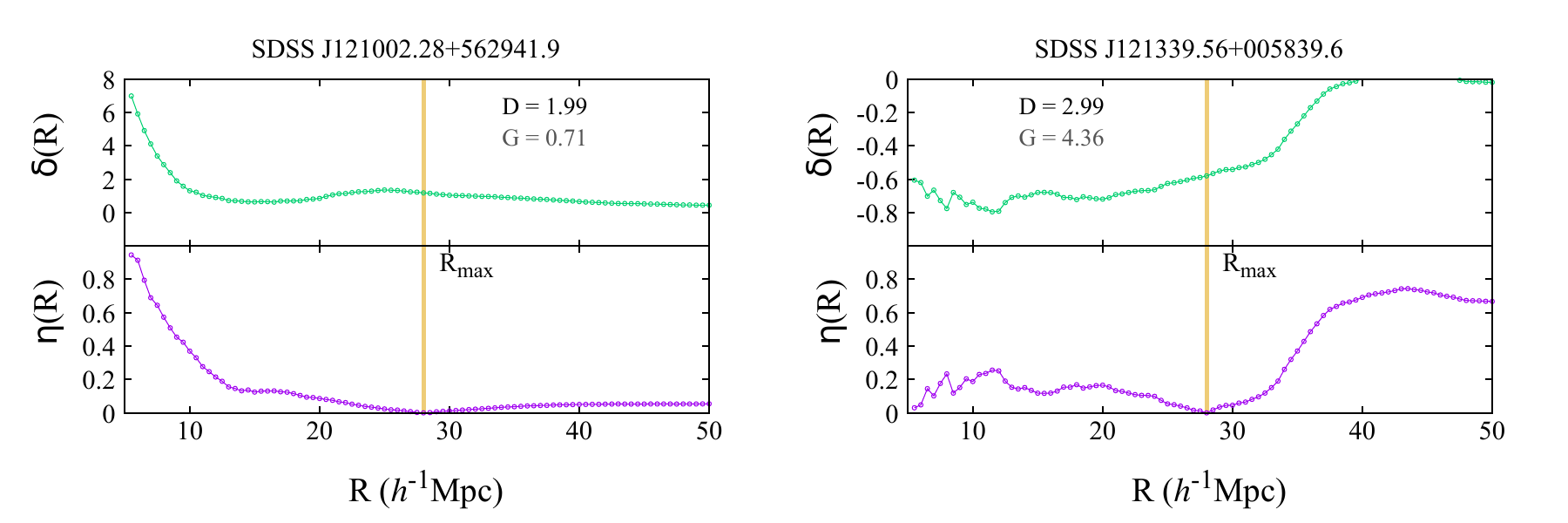}}}
\caption{ Variation of density contrast $\delta(R)$ and normalized absolute deviation in density contrast $\eta(R)$  with radius $R$, for two different galaxies with similar $R_{max}$ but lying in two different environments. \textbf{Based on the fitted values of $D$ and the observed variation in $\delta(R)$, the galaxy in the \textbf{Left Panel} is expected to lie in cluster-like environments, whereas, the galaxy in the \textbf{Right Panel} should be a part of a sheet or void}.}
\label{fig:delta_vary}
\end{figure*}

\noindent $\eta(R)$ can vary between 0 and 1. Now, $R_{max}$ is chosen to be the radii where $\eta(R) = 0$ is reached for the first time while increasing $R$. \autoref{fig:delta_vary} illustrates how the variation in $\eta(R)$ is used to identify $R_{max}$. Two different cases with similar $R_{max}$ are shown: [Left] A galaxy in a cluster, and [Right] a field galaxy.

\subsection{Local density}
To get the local density around any given galaxy we find the $k^{th}$ nearest neighbour distance, $ {\cal R}_{k}$ \cite{casertano85, verley07,sarkar20} for each galaxy and use it to find the local density ($\rho_{k}$) in the following way.

\begin{eqnarray}
\rho_{k} = \frac{3(k-1)}{4 \pi {\cal R}_{k}^3}
\label{eq:den_nn}
\end{eqnarray}

In this analysis, we choose $k=10$ for estimating the density of the galaxies. 

\subsection{Local geometric index}
The local dimension  or local density estimators, introduced in the literature earlier, have their limitations in uniquely quantifying the environment of a galaxy. The local density estimator can segregate the overdense and underdense regions admirably, however, it fails to inspect the geometrical nature of the environment. Two arbitrary regions in space can have very similar densities on an average but a completely different geometrical nature. The local dimension, on the other hand, is effective for quantifying the geometry of the galaxy distribution around the subject. However, it is unable to take into account information about the density at the location of the galaxy in consideration. Hence, it cannot tell the difference between sheets and curly filaments, or between voids and clusters. Also, the filaments observed on a larger length scale are found to be part of sheets. So it is difficult to uniquely determine whether a galaxy is lying in a filament or a sheet by only using local dimension.\\

\noindent We propose the \textit{local geometric index} estimator, which provides a relatively better classification of the local environment of a galaxy. The local geometric index ($G$) is defined as
\begin{eqnarray}
G &=&  D \,\, \left[ \, 1+ \, \beta \left( \, \log_{\bar{\rho}} \,\, \rho_{k} -1 \, \right) \right]. 
\label{eq:gld}
\end{eqnarray}

\noindent This empirical relation combines information from two quantities $D$ and $\rho$. This should theoretically be able to characterize the geometrical nature of the environment of galaxies, or any arbitrary point in a point distribution. After using an adaptive $R_{max}$, most of the cluster candidates are found around $D\sim 1.5$ but they are still mixed with galaxies with filament and sheets which lie between $D \sim 1$ and $D \sim 3$. The best way to segregate them is by using the local densities measured around these galaxies. For a moderately dense or sparse galaxy distribution ($<1 \hmpc$), the term inside the parentheses (\autoref{eq:gld}) would be $<0$ for overdense regions and $>0$ for underdense regions. If we consider the scaling factor $\beta=1$, the entire term inside the square brackets typically ranges between $0$ and $2$ for mean background density $\sim 10^{-2}\hmpc$. One can use smaller values of $\beta$ for distributions with higher mean densities. Also for a distribution with mean number density $>1 \hmpc$, $\beta$ has to be $-ve$. For an arbitrary distribution, a suitable choice for $\beta$ will be around $ - \frac{1}{2} \log{\bar{\rho}}$. This is to constrain the square bracket term within the range ($\geq0$, $\leq2$) and $G$ within ($\geq0$, $\leq6$) for most of the galaxies. However, for the case of individual galaxies found inside cluster cores or void centres, $G$ is expected to be $<0$ or $>6$.\\

The term in the parenthesis will scale up the local dimension of the galaxies which are in underdense parts of the cosmic web, whereas, for those lying in overdense regions, the local dimension will be scaled down. For the choice of $\beta=1$ that we have here, \autoref{eq:gld} reduces to $G=D \, \log_{\bar{\rho}} \, \rho_{k} $.  Being the product of the two factors, $G$ lies between $0$ and $6$. Galaxies residing in the clusters and voids will be pushed towards the two extremes $G \leq 0$ and $G \geq 6$ respectively. After the scaling of G, clusters are segregated from the filaments, but most importantly, the filaments having a higher density than that of the sheets are now segregated, slightly shifting towards the smaller 
values of $G$, $G < 3$ in particular. The field galaxies however get a higher value of $G$ and lie in the $G\geq 3$ region. We classify the galaxies into four main categories depending on their $G$ values.  The galaxies with $G < 1.5 $ are considered to be in clusters, whereas, $1.5 \leq G < 3 $ and  $3 \leq G < 4.5 $ represent galaxies which are part of filaments and sheets, respectively. The galaxies in voids are expected to have the highest values of $G$, so we consider those with $G \geq 4.5 $ to be lying in voids. This classification scheme is presented in \autoref{tab:class_env}. The performance of this classification is discussed in the results section. 

\begin{table*}{}
\centering
\caption{Local geometric index $G$ in different environments of the cosmic web}
\label{tab:class_env}
\begin{tabular}{|c|c|c|c|c|}
\hline
\textbf{Geometric Environment}     & Cluster & Filament & Sheet & Void \\
\hline
\textbf{local geometric index}     & $G < 1.5 $ & $1.5 \leq G < 3 $ & $3 \leq G < 4.5 $ & $G \geq 4.5 $ \\
\hline
\end{tabular}
\begin {itemize}
\item [] \scriptsize{}
\end{itemize}
\end{table*}

\subsection{Kolmogorov-Smirnov (K-S) Test}
\noindent In the next step, we try to check if there is a statistically significant difference between the local environment of two representative galaxies from different spiral galaxy samples. We perform a Kolmogorov-Smirnov (K-S) test for the distributions of $G$. Through K-S test, the supremum distance between two cumulative probability distributions for $G$ is obtained as

\begin{eqnarray}
\Delta_{KS} & = &  \mathrm{sup} \, \, \{ \,\, f_{m}(G)-g_{m}(G) \,\, \}
\label{eq:Dks}
\end{eqnarray}

\noindent Here $\rm{sup}$ is the supremum operator that finds the  maximum absolute difference. $f_{m}(G)$ and $g_{m}(G)$ are the cumulative probabilities for the two samples at $m^{th}$ checkpoint. If the effective number of data points in the two distributions are $n_1$ \& $n_2$ respectively, we will have $m=n_1+n_2$. \\

\noindent For a given significance level $\alpha$, the criteria for rejecting the null hypothesis (the two distributions are 
drawn from the same sample) is $\Delta_{ks} > \Delta_c ( \alpha )$. Here the critical value $\Delta_c ( \alpha )$ can be obtained from 

\begin{eqnarray}
\Delta_c ( \alpha ) = \sqrt{\, \, \left(\frac {1}{n_1}+\frac{1}{n_2}\right) \, \, \, \mathrm{ln} \, \left( \sqrt{\frac{2}{\alpha}}\right)}
\label{eq:pks}
\end{eqnarray}

\noindent  A sufficiently small $p$-value \footnote{ probability of accepting the null hypothsis } ( $p< \alpha$) for the obtained $\Delta_{KS}$ would lead to the rejection of the null hypothesis \cite{smirnov48}. The threshold value is usually 
set to $\alpha= 0.05$, which corresponds to $95 \%$ confidence level. 

%##############################################     RESULTS    #############################################################

\section{Results}
\subsection{Assesment of local geometric index}

\begin{figure*}
\centering
\resizebox{14 cm}{!}{\rotatebox{0}{\includegraphics{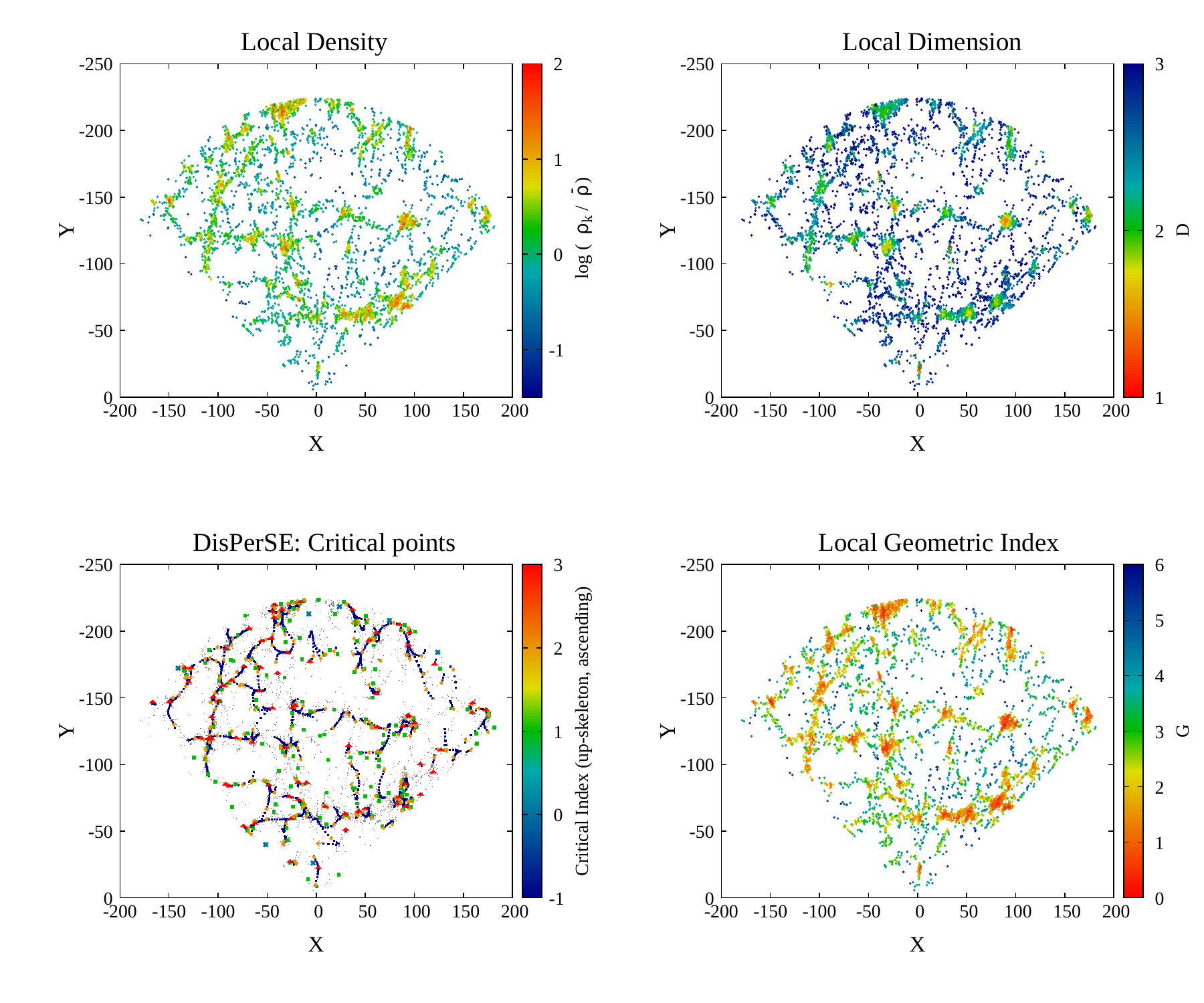}}}
\caption{A comparison of the environmental classifications of the background distribution, performed by using, local density \textbf{[Top left]}, local dimension \textbf{[Top right]}, DisPerSE \textbf{[Bottom left]} and local geometric index \textbf{[Bottom right]}. The colour bars show the respective quantities around the galaxies in the background galaxy sample. For the bottom-left panel, 
we have only shown the critical points of the 3-manifold for the background distribution (grey points), identified using DisPerSE \cite{sousbie08}. For each case, projection on the X-Y plane is shown 
for a $10 \hmpc$ slice on the Z-axis}
\label{fig:4_app}
\end{figure*}

\autoref{fig:4_app} shows the identification of the galaxy environments for the galaxies in the background sample (top left panel) using three different schemes apart from local geometric index. The top left and top right panels show the environment quantified using local density and local dimension (variable $R_{max}$) respectively. The right Panel at the bottom shows the critical points identified using DisPerSE \cite{sousbie08,sousbie08b} that characterize the 3-manifold for the background distribution. The indices of the critical points from which the manifold originates are shown for a persistence-ratio threshold of 3 with vertex-minima configuration. The obtained skeleton is smoothened 10 times. Maxima, minima or saddle points are shown with the ascending-order of manifolds. Indices 0,1,2 and 3 represent the voids, walls, filaments and nodes respectively. Points connecting the 1-manifold saddle points are shown with critical index -1. Lastly, the panel on the bottom right shows the environment of each of the galaxies in the background distribution identified by using the local geometric index. Unlike the local-density and local-dimension estimators, using the local geometric index we get sharp colour contrast for both large and moderate distances. Because it not only distinguishes between underdense and overdense regions but also categorizes galaxies based on the geometry of the cosmic web in which they are embedded. Comparing with the skeleton identified by DisPerSE, it is obvious that DisPerSE is able to trace the filaments and walls promisingly well, whereas the local geometric index is better at identifying genuine voids and clusters. It does not pick up any arbitrary junction or open-end of filaments as a cluster. It also identifies voids fairly well, even if they have tendrils passing through them. However, the local geometric index does not provide any sharp boundary segregating different cosmic web environments. \\

\begin{figure*}
\resizebox{7.5 cm}{!}{\rotatebox{0}{\includegraphics{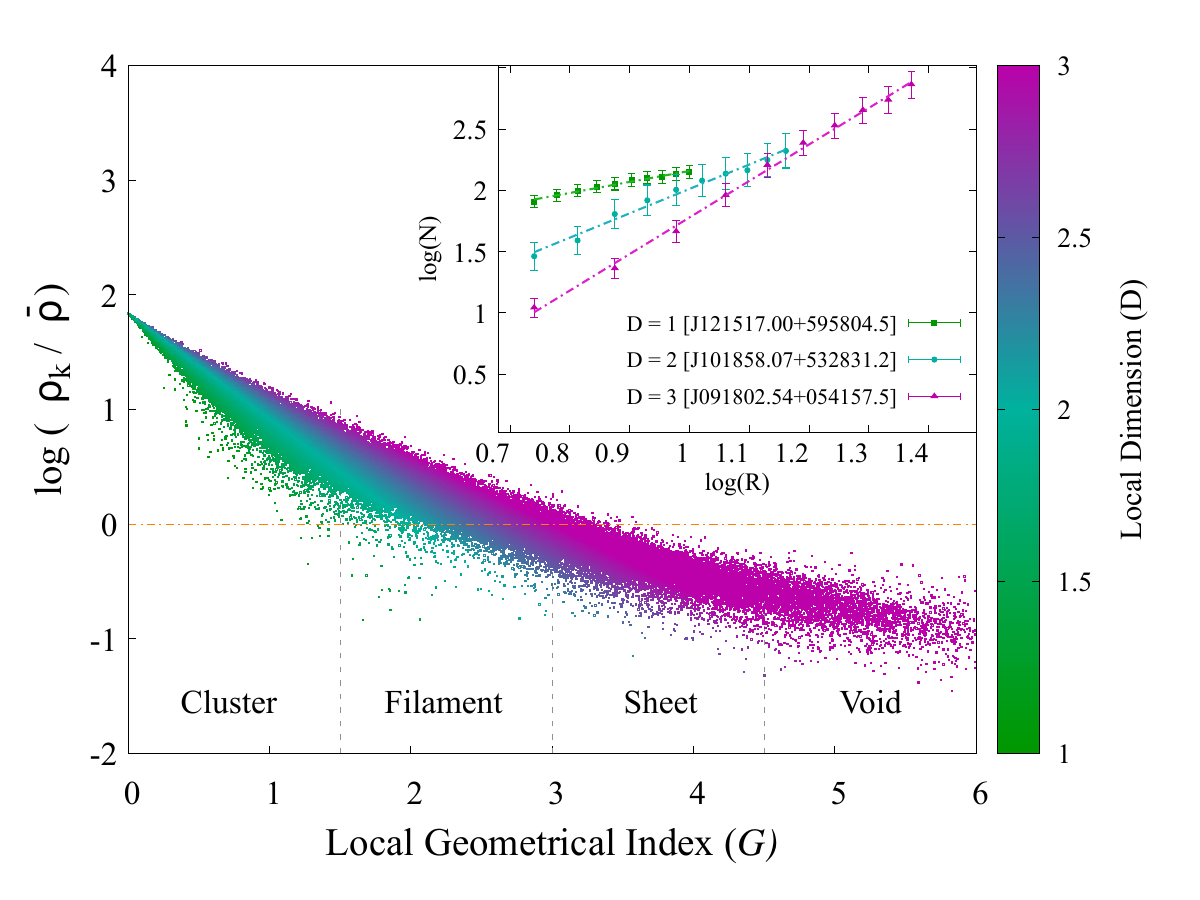}}} \hspace{5 px}
\resizebox{7.5 cm}{!}{\rotatebox{0}{\includegraphics{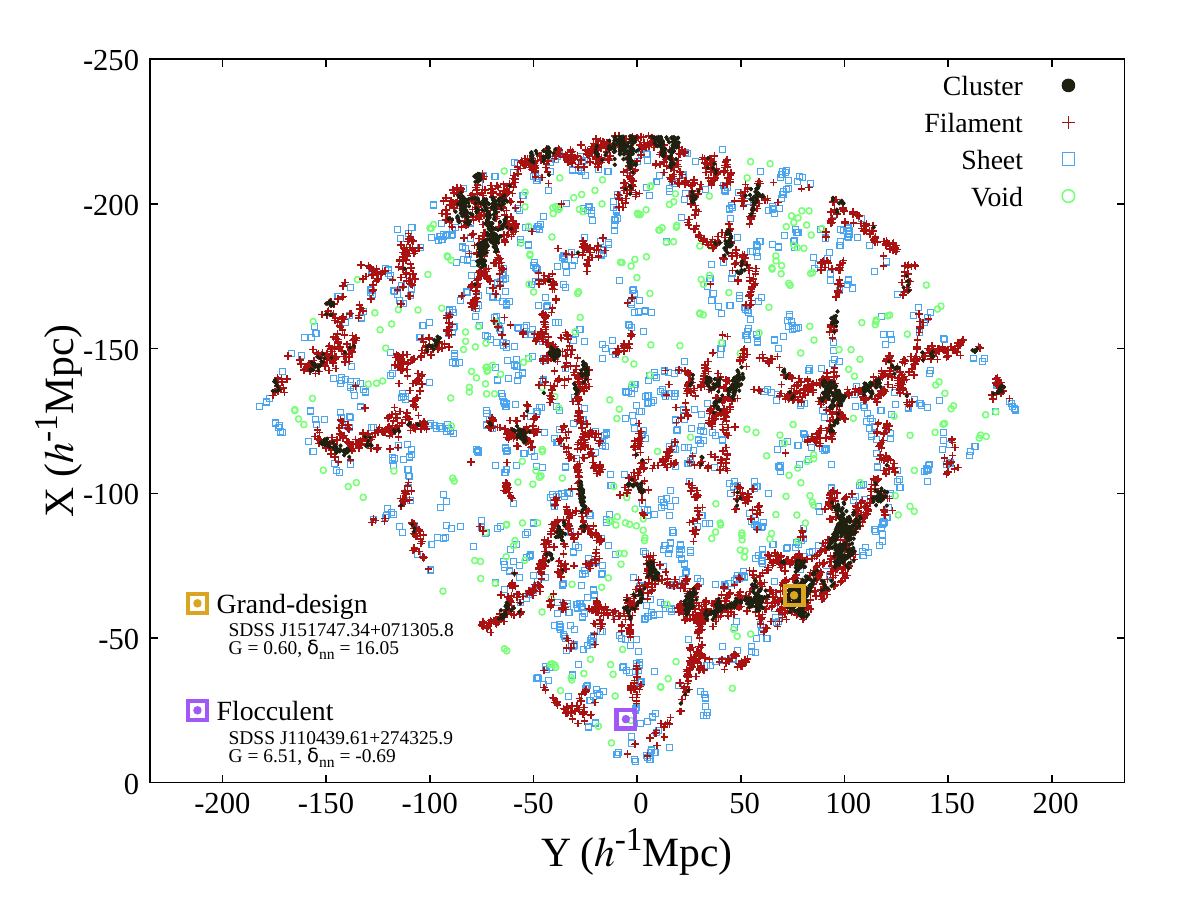}}}
\caption{\textbf{[Left]} This scatter plot shows the local geometric index $G$ along with the logarithmic overdensities ($\log{\rho_{k}}-\log{\bar{\rho}}$) around the galaxies in the background galaxy distribution. The local dimension $D$ is also shown with a variable colour palette. The least-square fitting for three different galaxies with three different values of $D$ is shown in the inset. \textbf{[Right]} The background galaxy distribution after classification, as projected on the X-Y plane for a $10 \hmpc$ slice on the Z-axis. A grand-design and a flocculent from the sample galaxies have been marked on the same.}
\label{fig:dendim}
\end{figure*}

\noindent The left panel of \autoref{fig:dendim} shows the profile of $G$ for the galaxies in the background galaxy distribution (\S 3.1), along with the logarithmic overdensities ($\log{\rho_{k}}-\log{\bar{\rho}}$) around the galaxy. The estimated local dimensions ($D$) are also presented by a variable colour palette. The least-square fitting for the local dimension estimation is demonstrated in the inset of \autoref{fig:dendim}, for three different galaxies with three different values of $D$. The right panel of \autoref{fig:dendim} shows the galaxies in the background galaxy distribution, categorized into four different types, based on the values of $G$. Following the segregation scheme that we have tabulated in \autoref{tab:class_env}, we get the majority of the galaxies in the background sample ($\sim 78 \%$) in filaments and clusters. The rest of the galaxies are part of underdense regions like sheets and voids (\autoref{tab:env_num}).The average density contrast for the galaxies in four types of environments is found to be $10.64$, $1.05$, $-0.49$, and $-0.81$ respectively. Though different samples and density estimation methods are used, \cite{tsaprazi22} exhibits a similar trend of density contrast in the various environments for SDSS galaxies. We confirm that the local geometric index successfully characterizes the environment by comparing the population fractions with earlier works in literature \cite{tempel14, sarkar19, malavasi20, tsaprazi22}, and through visual assessment of the classified distributions. We are unable to analyse the environments of 3 grand-designs and 10 flocculents. For these galaxies, local dimension estimation does not meet our chosen criteria for chi-square minimization ($\frac{\chi^2}{\nu}\leq 0.5$). Hence, we finally have the information about the environment of $367$ grand-designs and $619$ flocculents.

\subsection{Environment of grand-designs and flocculents }
\begin{figure*}
\centering
\hspace{-5 px} \resizebox{7.5 cm}{!}{\rotatebox{0}{\includegraphics{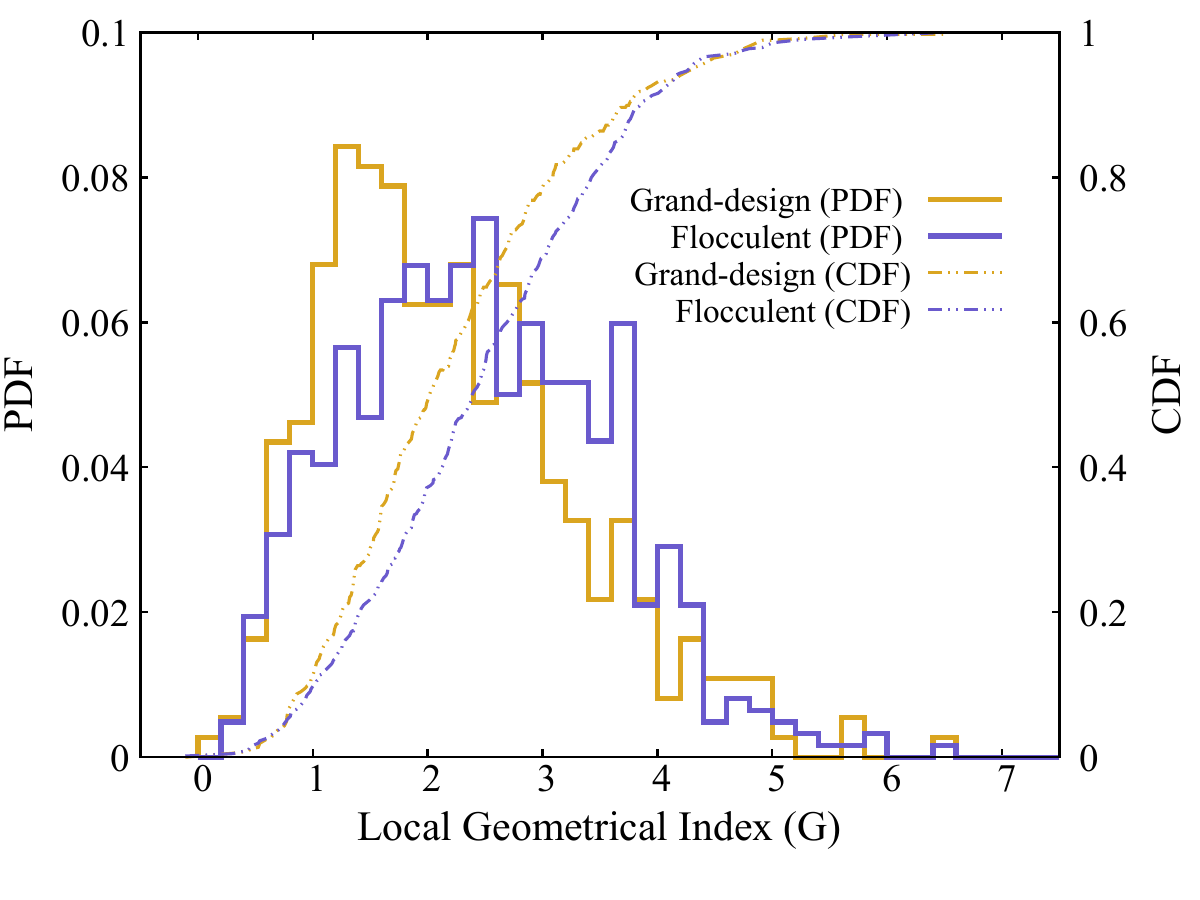}}} \hspace{6 px}
 \resizebox{7.5 cm}{!}{\rotatebox{0}{\includegraphics{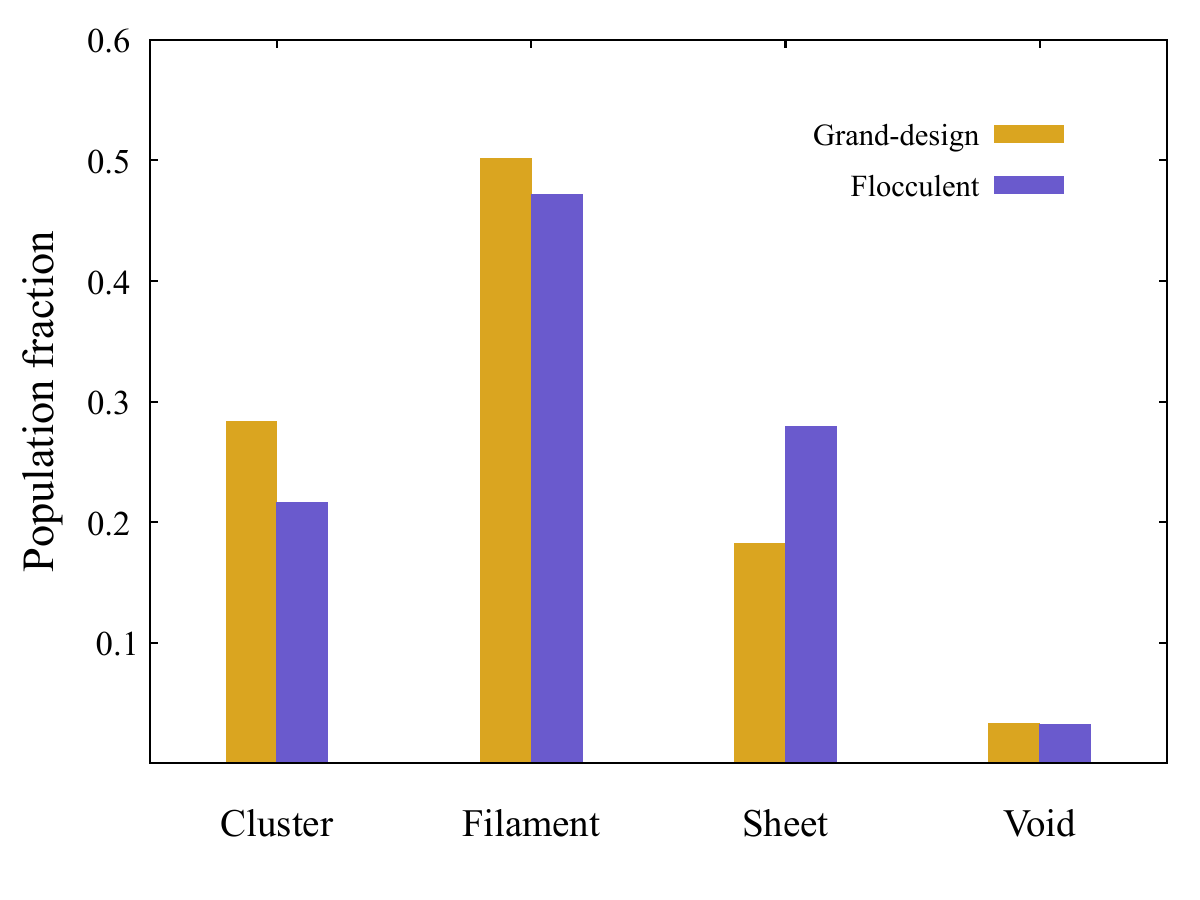}}}\\
 \caption{\textbf{[Left]} Probability distribution and Cumulative distribution function of grand-designs and flocculents with respect to the local geometric index $G$; \textbf{[Right]} Fraction of population of grand-designs and flocculents in different environments.}
 \label{fig:results1}
 \end{figure*} 

\begin{table*}{}
\centering
\caption{Galaxy count for our sample spirals and all background galaxies in different environments of the cosmic web}
\label{tab:env_num}
\begin{tabular}{|l|c|c|c|}
\hline
Galaxies                    & \textbf{ grand-designs}     & \textbf{flocculents}        & Background        \\
                \hline
\textbf{in clusters}        & 104 \textit{(28.34 \%)}     &  134 \textit{(21.65 \%)}    & 29901 \textit{(32.90 \%)} \\
\textbf{in filaments}       & 184 \textit{(50.14 \%)}     &  292 \textit{(47.17 \%)}    & 41327 \textit{(45.47 \%)} \\
\textbf{in sheets}          &  67 \textit{(18.26 \%)}     &  173 \textit{(27.95 \%)}    & 16067 \textit{(17.68 \%)} \\
\textbf{in voids }          &  12 \textit{( 3.27 \%)}     &   20 \textit{( 3.23 \%)}    &  3587 \textit{( 3.95 \%)} \\
\textbf{Total }             & \textbf{367}                & \textbf{619}                & 90882            \\
\hline
\end{tabular}
\begin {itemize}
\item [] \scriptsize{}
\end{itemize}
\end{table*}

\noindent In \autoref{fig:results1} [Left Panel], we compare the cumulative distribution function ($CDF$) as well as the probability distribution function ($PDF$) of {$G$} for the grand-designs and flocculents in our data sets. We present the distribution of  grand-designs and flocculents in different local geometric environments in \autoref{fig:results1} [Right Panel]. The four web environments namely voids, sheets, filaments, and clusters are characterized by choosing different ranges of $G$ (\autoref{tab:class_env}). In \autoref{tab:env_num} we present the results showing the fraction of grand-designs and flocculents in these four environments. Being the densest regions, clusters and filaments in general host a major fraction of all galaxies in the Universe. The overall fraction of galaxies is expected to be higher in filaments. We note that more than $78\%$ of the grand-designs reside in clusters and filaments, whereas the fraction of flocculents in sheets and voids is $31\%$, $\sim 10\%$ higher than the grand-designs. Therefore, from the primary observation, it is clear that the grand-designs are mostly located in dense environments like clusters and filaments, whereas the flocculents are more abundant in sparser components of the cosmic web, like the sheets and the voids. \\

\noindent Next, we check if this observed difference is statistically significant, using a K-S test. In \autoref{fig:results1}, the supremum difference between the $CDF$s of our sample grand-designs and the flocculents is found to be $0.142$. \autoref{tab:ks} shows that the critical values to reject the null hypothesis i.e., both samples are drawn from the same distribution with 90 \%, 99 \%, and 99.9 \% confidence are $0.081$, $0.107$ and $0.128$ respectively. We note that all values are smaller than the obtained supremum value. Therefore, we can now statistically confirm that the grand-designs dwell in a substantially different environment than the flocculents with $> 99.9\%$ confidence. 
Our results are in compliance with earlier observations that the fraction of grand-designs is correlated with the local background density \cite{ann14}.\\

\begin{table*}{}
\centering
\caption{K-S tests for the distribution of the local geometric index $G$ for our sample spirals}
\label{tab:ks}
\begin{tabular}{|c|c|c|c|c|c|c|c|c|c|}
\hline
\multicolumn{2}{|c|}{grand-design} & \multicolumn{2}{|c|}{flocculent} & \multirow{2}{*}{$\Delta_{KS}$} & \multicolumn{3}{|c|}{ $\Delta_{c} $} & \multirow{2}{*}{$p$ - value} \\
  \cline{1-4}  \cline{6-8} 
count ($n_1$) & median $G$ & count ($n_2$) & median $G$ & & 90 \% & 99 \% & 99.9 \% & \\
\hline 
\rule{0pt}{25 pt} $367$ & $2.01^{+0.84}_{-0.65}$ & $619$ & $2.39^{+0.86}_{-0.75}$ & \textbf{0.142} & $0.081$ &  $0.107$ & $0.128$ & $1.57 \times 10^{-4}$\\
&&&&&&&& \\
\hline
\end{tabular}
\begin {itemize}
\item [] \scriptsize{$n_1$ and $n_2$ are the effective number of galaxies in the sample, and its counterpart, respectively. The critical K-S distance for different confidence levels ($\Delta_c$) is provided along with the measured K-S distance ($\Delta_{KS}$) for comparison.}
\end{itemize}
\end{table*}

\subsection{Physical properties of galaxies in different environment}
\begin{figure*}
\centering
\hspace{-20 px}
\resizebox{16 cm}{!}{\rotatebox{0}{\includegraphics{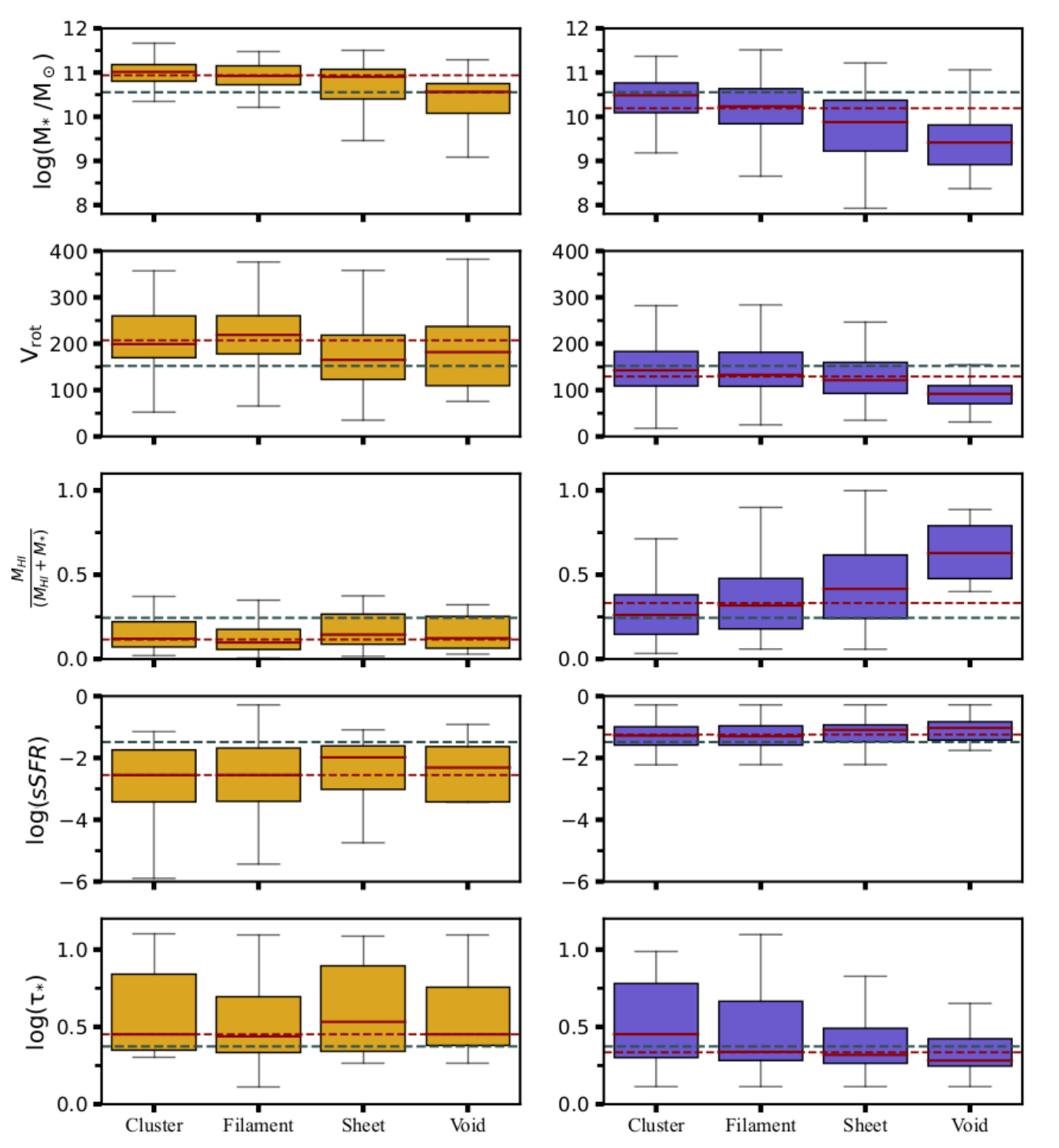}}} 
\caption{This figure presents a comparison between the physical properties of grand-designs and flocculents in different environments \textbf{[Left]} grand-designs (in golden-rod); \textbf{[Right]} flocculents (in slate-blue). The median values ($Q_2$) are shown by the red lines; the lower and upper end of the boxes show the 1st and 3rd quartiles ($Q_1$ and $Q_3$). The bottom and top edges of the error bars indicate $Q_0$ and $Q_4$. The teal dashed line shows the global median for all $986$ spirals. The median values for each class are shown by the red dashed lines. \textbf{[Top]} Distribution of log-stellar mass in dex; \textbf{[2nd from top]} asymptotic rotational velocity; \textbf{[3rd from top]} HI Mass fraction; \textbf{[4th from top]} log-sSFR in dex; \textbf{[Bottom]} Age of the stellar population. \textbf{Units:} $V_{rot}-\mathrm{Km\;s^{-1}}$; $sSFR - \mathrm{Gyr^{-1}}$; $\tau_{\star}- \mathrm{Gyr}$.}
\label{fig:props_env}
\end{figure*}
 
\noindent  A detailed overall comparison of the physical properties of grand-designs and flocculents is presented by \cite{gdfl2022}, using their asymptotic rotational velocity ($V_{\rm{rot}}$), the fraction of the neutral hydrogen mass-to-blue luminosity ($M_{HI}/L_{B}$) and de Vaucouleurs class indices ($T$). It is found that, on average, the grand-designs have a comparatively higher dynamical mass than the flocculents. Besides, the sample of flocculents is comprised of late-type galaxies with higher gas content. In this paper, we study the stellar mass ($M_{\star}$), specific star formation rate ($sSFR$), the age of the stellar population ($\tau_{\star}$) and the gas fraction ($M_{HI}/(M_{\star}+M_{HI})$) along with $V_{\rm{rot}}$ for the grand-designs and flocculents, separately in different geometric environments. \\

\noindent In \autoref{fig:props_env}, we use box plots to present the distribution of physical properties of the two classes of spirals lying in different environments. The median values ($Q_2$) for each environment are shown with the red solid lines. The 1st and 3rd quartiles ($Q_1$ and $Q_3$) are represented by the lower and upper edges of the boxes. The top and bottom of the error bars indicate $Q_4$ and $Q_0$. The teal dashed line shows the global median for the combined sample. The median values for each class are separately shown by the red dashed lines. Both for flocculents and grand-designs. The median values of all the properties in different environments are tabulated in \autoref{tab:PhyProps}. \\

\noindent The top panel (\autoref{fig:props_env}) shows the distribution of stellar masses. Note that the stellar mass for both types of galaxies systematically increases as we move from voids to clusters; also, the grand-designs in any environment are more massive than their flocculent counterparts. In fact, the grand-designs in clusters and filaments are found to be more than 10 times more massive than the flocculents found in fields (\autoref{tab:PhyProps}). \\

\noindent The actual mass of a galaxy also includes the gas mass and the mass of the dark matter component. The mass of the host dark matter halo is believed to be the most dominating factor in the evolution of galaxies \cite{white78,cole00,springel05,contreras15}. Although the spiral activity clearly depends on the mass of the galaxy, it is evident here that not only the mass but the environment also plays a crucial role in regulating the nature of the spiral arms. For the galaxies in our sample, we do not have direct measurements of the dark matter halo mass. The stellar and gas masses may only give a rough estimate of the overall mass of the galaxy. The asymptotic rotational velocity of the galaxies   $V_{\rm{rot}}$, on the other hand, is a proxy for the total dynamical mass of the galaxy. \noindent In the second panel from the top (\autoref{fig:props_env}) we show $V_{\rm{rot}}$ of grand-designs [Left] and flocculents [Right] in different environments. We note that of all different categories, the grand-designs in filaments have the highest $V_{\rm{rot}}$ and hence higher dynamical masses. The grand-designs in the filaments are found to be the most abundant in dark matter \autoref{tab:PhyProps}), whereas, the flocculents in fields have the least. The filaments are the final routes of mass flow from the voids into the clusters. As a result, due to the unidirectional flow of mass, filaments exhibit a higher degree of momentum transfer than other environments. Thus, massive galaxies near the filament spine would experience larger tidal torques. \cite{codis12} shows that the spin angular momentum of massive dark-matter halos has a higher probability of being perpendicular to the direction of the nearby filament. These halos exhibit a significantly larger angular momentum \cite{lopez21} compared to the less massive halos having their spin aligned with the direction of the filaments. As a result, the angular momentum of galaxies with large dynamical masses (especially grand-designs) is more likely to be boosted in filaments.\\

\begin{table*}{}
\centering
\caption{Median values of the physical properties for grand-designs and flocculents.}
\label{tab:PhyProps}
\begin{tabular}{|c|c|c|c|c|c|c|c|c|}
\hline
\multirow{2}{*}{Properties}&\multicolumn{4}{|c|}{grand-design} & \multicolumn{4}{|c|}{flocculent}  \\
  \cline{2-4}  \cline{5-9} 
	&	Cluster 	&	Filament 	&	Sheet 	&	Void	&	Cluster 	&	Filament 	&	Sheet 	&	Void	\\
 \hline
$\rm{\log(M_{\star}/m_\odot)}$	&	11	&	10.9	&	10.9	&	10.6	&	10.5	&	10.2	&	9.9	&	9.4	\\
$\rm{V_{rot}}$	&	199.2	&	218.9	&	165.2	&	181.9	&	142.5	&	132.9	&	121.3	&	92	\\
$\rm{M_{HI}/(M_{HI}+M_{\star})}$&	0.12	&	0.1	&	0.14	&	0.12	&	0.26	&	0.32	&	0.42	&	0.63	\\
$\rm{\log(sSFR)}$	&	-2.55	&	-2.55	&	-1.98	&	-2.31	&	-1.3	&	-1.3	&	-1.1	&	-1	\\
$\rm{\log(\tau_{\star})}$	&	2.83	&	2.74	&	3.41	&	2.83	&	2.8	&	2.2	&	2.1	&	1.9	\\
\hline
\end{tabular}
\begin {itemize}
\item [] \scriptsize{\textbf{Units:} $V_{rot}-\mathrm{Km\;s^{-1}}$; $sSFR - \mathrm{Gyr^{-1}}$; $\tau_{\star}- \mathrm{Gyr}$.}
\end{itemize}
\end{table*}

\noindent \noindent In the third panel from the top (\autoref{fig:props_env}), we analyze the abundance of gas in our sample galaxies in different environments. We find that the median $M_{HI}/(M_{\star}+M_{HI})$ systematically increases from clusters to voids for the flocculents. However, for the grand-designs, there is no such trend. In all environments, the flocculents are found to be richer in gas. Especially, the flocculents in fields are 4-5 times higher in gas content compared to an average grand-design (\autoref{tab:PhyProps}). This confirms that the cold gas content in a galaxy is crucial in the manifestation of its flocculent nature, and persistence of the spiral activity \cite{SellwoodCarlberg1984}. Local non-axisymmetric instabilities could grow faster in a cold gas disc because of its lower disc stability. Also, a colder component decreases the radial group velocity of the propagating instability \cite{Ghosh2015} and hence increases the lifetime of the spiral activity. Simulations by \cite{Elmegreen1993} show that the flocculent spirals are formed in the colder gas component which is more abundant in our sample of flocculents. \\ \\

\noindent In the fourth panel from the top (\autoref{fig:props_env}), we show the specific star formation rate. In all environments, the flocculents show a much higher star formation rate compared to the grand-designs. Besides, both grand-designs and flocculents in sheets and voids exhibit an overall higher star formation rate compared to their counterparts in clusters and filaments. The flocculents in voids are found to be the highest $sSFR$ among all. Therefore, it is well-known that, unlike clusters, voids and sheets don't trigger the quenching mechanism in flocculent spirals and allow stochastic star formation \cite{Gerola1978}. The flocculent nature of spiral galaxies is boosted by this kind of propagation of star formation bursts. This is a possible reason for flocculents to be found more in sheets and voids. In the two classes of spirals in different environments, we also look for the age of the stellar population ($\tau_\star$). This is to find any possible links between the local environment of the grand-designs and flocculents and the merger history of the dark matter halos in which they reside. A halo that is formed much earlier is expected to host an older stellar population in general. \\ \

\noindent In the bottom panel of \autoref{fig:props_env} we show that the flocculents overall have a stellar population that is much younger compared to the grand-designs, especially in voids and sheets. These young low-mass galaxies have a large gas reservoir, making them prime candidates to experience frequent starbursts and stochastic star formation. On the other hand, the grand-designs in clusters and filament are 
populated with a much older population, indicating possible halo and galaxy mergers. A recent study by \cite{zheng17} states that the environment of a galaxy is rather weakly related to its stellar age. On the contrary, \cite{tiwari20} reports the galaxies in voids are younger compared to their counterparts in clusters and filaments. We observe a similar trend for the spirals in our sample. We find that the grand-designs in sheets host an older stellar population, even compared to filaments. Interestingly, these galaxies also have a higher gas content and significantly greater star formation rate compared to the grand-designs in other environments (\autoref{tab:PhyProps}). The mere fact that these grand-designs have remained actively star-forming for quite a while indicates that they must have had a sufficient and steady supply of gas, in order for the spiral activity to grow and persist without the quenching of star formation.\\
 
%##############################################     Discussion    #############################################################

\section{Discussion}

\begin{itemize}
    \item We consider the galaxy of interest (grand-design or flocculent) at its position along with the background galaxy distribution, which is a magnitude-limited sample and serves as an underlying framework. Any distribution of galaxies without a considerable Malmquist bias or sampling bias is suitable for the estimation of $G$. \\
    
    \item We make necessary corrections for the galaxies at the boundary while calculating $G$. The count in a partially accessible 
    sphere is scaled using the volume fraction to get the probable galaxy count in a complete sphere. \\

    \item The definition of local dimension given here is somewhat similar to that of fractal dimension \cite{mandelbrot83}. However, the method of their estimation and their purpose differ from each other. The fractal dimension is used to investigate the self-similarities in the structural form of the test object on different length scales. On the other hand, the local dimension on a given length scale quantifies the local geometric environment of a galaxy embedded in a larger structure.
    
    \item We note that our analysis is partly affected by redshift space distortions (RSD). Some spurious filament-like structures may appear on small scales due to the {\it Finger of God} effect, elongating the virial clusters in redshift space. In \autoref{sec:rsdmill}, we show the effect of RSD on the measurement of $G$ using N-body simulation data from Millenium simulation. We find the population fraction of galaxies in filaments and clusters effectively changes due to RSD. However, the focus here is not to identify individual filaments or clusters, but to statistically test how different the geometric environments of the grand-designs and flocculents are. As both of the samples are analysed in redshift space, the results presented here despite RSD provide sufficient evidence to infer that the grand-designs and flocculents do not prefer the same geometric habitats.\\

    \item Different population fractions of galaxies in the four cosmic web environments have apparently been noted in studies to date. Here, we do not even attempt to compare the fraction to any of the previous studies. One can determine a different set of $G$ values shown in the \autoref{tab:class_env} and get different population fractions. We select the threshold values simply at equal spacing, relying more on visual confirmation.\\

    \item There are different ways to classify the morphology of spiral galaxies. One can classify them into their popular Hubble or de-Vaucouleur classes. However, one can also classify spiral galaxies just based on the presence or absence or type of bulges and/or bars, and/or spiral arms. Likewise, galaxies can be classified based on the two types of spiral arms: the grand-designs and the flocculents. The intermediate class of galaxies are called multiple-arms. To do a reliable statistical analysis, we need access to the necessary information for a sufficient number of multiple-arm galaxies, which is not available to us. In this paper, our goal is to find out if the local environment determines the type of spiral arms present in a galaxy, restricting ourselves to grand-design and flocculent types of spiral arms.

    \item $\gamma$ varies from galaxy to galaxy as it incorporates the information of density as well as the geometry of the cosmic web environment. In this analysis, $\gamma$ for the sample galaxies, takes values within the range -2 to +2.  In most cases, galaxies in clusters have $+ve$ $\gamma$ and, for voids, $\gamma$ is found to be $-ve$.

\end{itemize}

%##############################################     Conclusion   #############################################################

\section{Conclusion}

 We introduce a new estimator {\it local geometric index ($G$)} as a hybrid of local dimension and local density, to investigate the geometrical environment of the grand-design and flocculent spiral galaxies. The cosmic web environment, in the vicinity of a total of $986$ spiral galaxies, is studied and quantified. The galaxies are segregated into four different types based on their local geometric index. We find that the grand-designs are massive spirals with relatively older stellar populations, which are found mostly in the environments where tidal forces are prevalent. The flocculents, on the other hand, are gas-rich, late-type spirals mostly found in isolated environments, where tidal fields are weak. We also observe that the grand-designs in the fields have a notably older but actively star-forming population. Overall, our study shows that the two main reasons that are responsible for the formation and persistence of spiral activity are tidal torques and cold gas abundance. We plan to carry out a detailed study with a larger set of grand-design and flocculents, along with the intermediate multiple-arm type galaxies, to investigate the problem in greater detail.

\section*{Data-availability}
The data utilized in this study can be found in Hyperleda and databases of SDSS and  Millennium Run Simulation, which are open to the public. Data generated in this work can be shared on request to the corresponding author.

\section*{Acknowledgement}
The authors acknowledge and appreciate the insightful remarks and recommendations of the anonymous reviewer. SS thanks DST-SERB for support though the National Post-Doctoral Fellowship (PDF/2022/000149). SS also thanks Prof. Somnath Bharadwaj for useful discussions and suggestions. Funding for the Sloan Digital Sky Survey IV has been provided by the Alfred P. Sloan Foundation, the U.S. Department of Energy Office of Science, and the Participating Institutions. SDSS-IV acknowledges support and resources from the Center for High Performance Computing  at the University of Utah. The SDSS website is www.sdss.org. SDSS-IV is managed by the Astrophysical Research Consortium for the Participating Institutions of the SDSS Collaboration including the Brazilian Participation Group, the Carnegie Institution for Science, Carnegie Mellon University, Center for Astrophysics | Harvard \& Smithsonian, the Chilean Participation Group, the French Participation Group, Instituto de Astrof\'isica de Canarias, The Johns Hopkins University, Kavli Institute for the Physics and Mathematics of the Universe (IPMU) / University of Tokyo, the Korean Participation Group, Lawrence Berkeley National Laboratory, Leibniz Institut f\"ur Astrophysik Potsdam (AIP),  Max-Planck-Institut f\"ur Astronomie (MPIA Heidelberg), Max-Planck-Institut f\"ur Astrophysik (MPA Garching), Max-Planck-Institut f\"ur Extraterrestrische Physik (MPE), National Astronomical Observatories of China, New Mexico State University, New York University, University of Notre Dame, Observat\'ario Nacional / MCTI, The Ohio State University, Pennsylvania State University, Shanghai Astronomical Observatory, United Kingrand-designom Participation Group, Universidad Nacional Aut\'onoma de M\'exico, University of Arizona, University of Colorado Boulder, University of Oxford, University of Portsmouth, University of Utah, University of Virginia, University of Washington, University of Wisconsin, Vanderbilt University, and Yale University.

\bibliographystyle{JHEP}
\bibliography{gdfl_env}
\appendix 
\section{ Effect of redshift space distortions in N-body Simulation }
\label{sec:rsdmill}
We use data from the semi-analytic galaxy catalogue \cite{henriques15}, based on the Millennium run simulation \cite{springel05}
use cosmological parameters from PLANCK first-year data to update the Munich model of galaxy formation. Galaxies at redshift 0, with the same absolute magnitude range as the background data (SDSS magnitude-limited sample, see \S3.1) are extracted using an 
SQL query. We map the position of the galaxies in the simulated data into redshift space using their peculiar velocities considering the observer at (0,0,0). Cubic regions of $200 \hmpc$ containing $80000$ randomly sampled galaxies are taken out from the entire volume, both from real space and redshift space distributions. We use $G$ to find the four different types of environments in the Millennium distribution, both in real and redshift space (\autoref{fig:rsd_mill}). We find that RSD effectively reduces the fraction of galaxies in clusters, resulting in an apparent increase in the fraction of galaxies found in filaments. Hence, our analysis and the population fraction of clusters and filaments presented in this work are partly affected by RSD.

\setcounter{figure}{0}    
\counterwithin{figure}{section}

\begin{figure*}
\resizebox{16 cm}{!}{\rotatebox{0}{\includegraphics{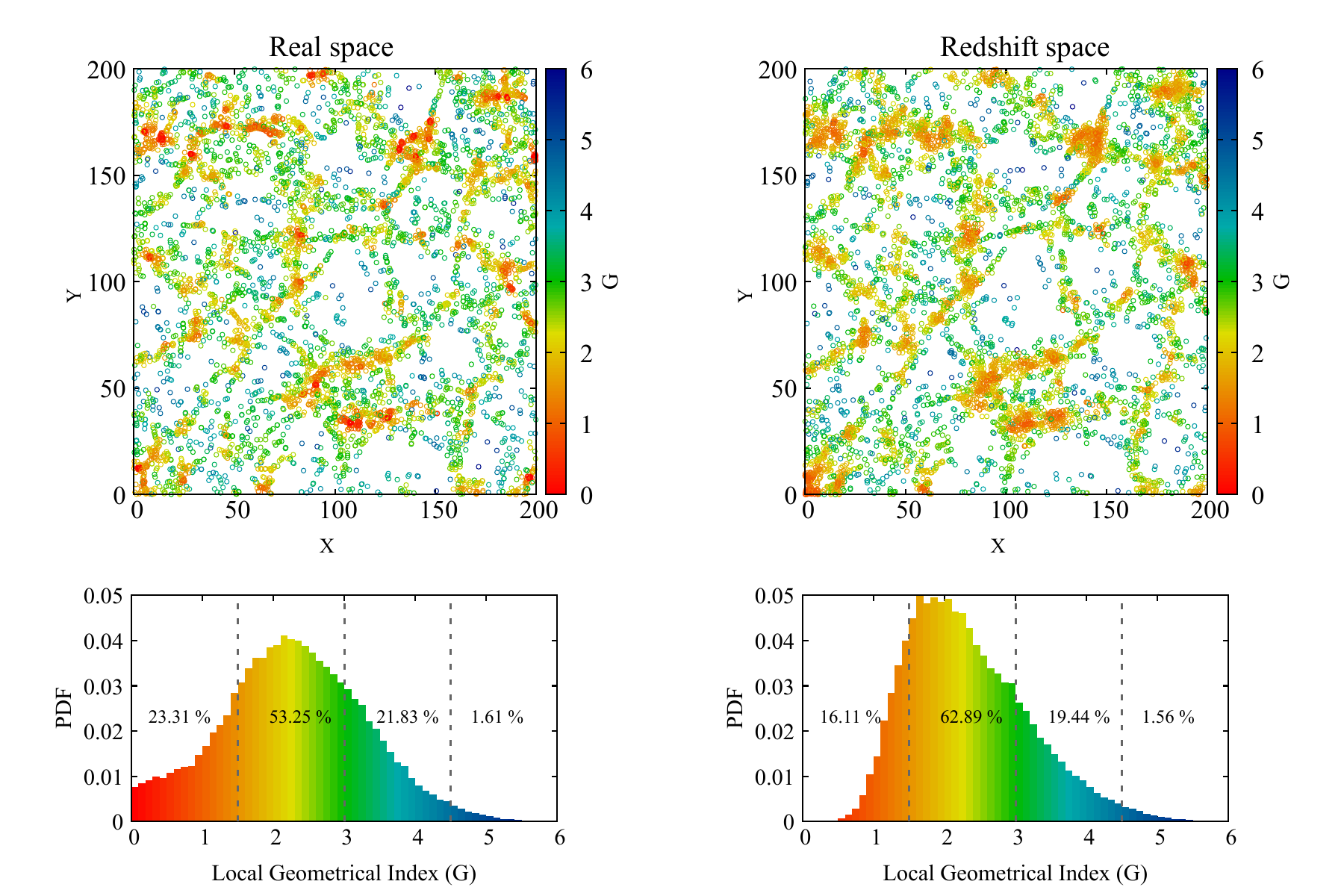}}} 
\label{fig:rsd_mill}
\caption{ This figure shows the Millennium galaxies in different environments, in real space \textbf{[left]} and redshift space \textbf{[right]}.
galaxies within a $20 \hmpc$ slice are projected on the X-Y plane. The fraction of galaxies in different environments is also listed in the legends.}

\end{figure*}

\label{lastpage}
\end{document}